# Optically Induced Ferromagnetic Order in a Ferrimagnet

Sergii Parchenko[1*], Agne Åberg Larsson[2,$], Vassilios Kapaklis[2], Sangeeta Sharma[3,4], Andreas Scherz[1]

1. European XFEL, Holzkoppel 4, 22869 Schenefeld, Germany
2. Department of Physics and Astronomy, Uppsala University, Box 516, 751 20 Uppsala, Sweden
3. Institut für Nichtlineare Optik und Kurzzeitspektroskopie, Max-Born-Straße 2A, 12489 Berlin, Germany.
4. Institute for theoretical solid-state physics, Freie Universität Berlin, Arnimallee 14, 14195 Berlin, Germany

*Corresponding authors: sergii.parchenko@xfel.eu

**The parallel or antiparallel arrangement of electron spins plays a pivotal role in determining the properties of a physical system. To meet the demands for innovative technological solutions, extensive efforts have been dedicated to exploring effective methods for controlling and manipulating this arrangement [1]. Among various techniques, ultrashort laser pulses have emerged as an exceptionally efficient tool to influence magnetic order. Ultrafast suppression of the magnetic order [2,3], all-optical magnetization switching [4, 5, 6, 7], and light-induced magnetic phase transitions [8] are just a few notable examples. However, the transient nature of light-induced changes in the magnetic state has been a significant limitation, hindering their practical implementation. In this study, we demonstrate that infrared ultrashort laser pulses can induce a *ferro*magnetic arrangement of magnetic moments in an amorphous TbCo alloy, a material that exhibits *ferri*magnetism under equilibrium conditions. Strikingly, the observed changes in the magnetic properties persist for significantly longer durations than any previously reported findings. Our results reveal that ultrashort optical pulses can generate materials with identical chemical composition and structural state but entirely distinct magnetic arrangements, leading to unique magnetic properties. This breakthrough discovery marks a new era in light-driven control of matter, offering the exciting potential to create materials with properties that were once considered unattainable.**

Over recent years, significant progress has been achieved in developing tools to program and control the quantum properties of materials [9, 10]. The strong coupling and competition between spin, orbital, charge, and lattice order create a situation where several different states may have similar energies or exist in a superposition [11]. These competing degrees of freedom give rise to various exotic states of matter, such as high-temperature superconductivity [12], van der Waals heterostructures [13], and diverse magnetic arrangements [14]. The electronic configurations of these materials are highly sensitive to external perturbations, particularly through the use of ultrashort optical pulses [15]. The spectral tunability of ultrashort light pulses allows for selective excitation of specific degrees of freedom, resulting in desired changes in coupled orders. Several remarkable studies have demonstrated light-induced magnetic phase

---

[$] Present Address: Ericsson Research, Torshamnsgatan 23, 16483 Stockholm, Sweden.

transitions [16, 17], enhanced superconducting correlations [18, 19], suppression of long-range magnetic order [20, 21], and the interplay between different orders following optical excitation [22]. However, the resulting changes are often short-lived, with the system returning to an equilibrium state within a few tenths or hundreds of picoseconds at most, posing challenges for their practical implementation.

The examples of ultrashort light pulse applications typically involve driving materials to known states without creating otherwise inaccessible states. However, there are instances where strong optical excitation can achieve unique configurations, like transient ferromagnetic-like states in GdFeCo rare-earth-transition metal alloys [23] or Floquet-Bloch state excitation [24]. These exotic states usually dissipate shortly after optical excitation. While long-lived light-induced changes have been observed in spin-crossover molecules [25] and some photoinduced anisotropy garnets [26], they often require prolonged light exposure. The primary challenge in achieving enduring, non-destructive modifications lies in recovery dynamics. To overcome this, inducing multiple processes through optical excitation can push the system into an energetic configuration with a formidable potential barrier compared to the equilibrium state, where $\Delta E \gg kT$, making thermal recovery improbable.

In this study, we demonstrate that the combined dynamics of optically induced spin transfer (OISTR)[27, 28, 29] and optically induced demagnetization processes [2, 3, 30] can significantly change the electronic state in TbCo alloys. This results in a situation where the system takes a significantly longer time to fully recover to an equilibrium state compared to when OISTR and demagnetization processes are disentangled in time. Specifically, we show that after excitation with a near-infrared (NIR) ultrashort laser pulse, the initially ferrimagnetic TbCo alloy undergoes a transition and becomes a ferromagnet. Importantly, the induced ferromagnetic state in TbCo is remarkably long-lived, potentially persisting for several tenths of a nanosecond or even up to hundreds of nanoseconds. This striking observation is a consequence of the enduring modification of the electron configuration in Co and Tb $d$ states, which alters the sign of the exchange interaction between the Tb and Co sublattices.

**Results.** RE-TM alloys, including TbCo, have garnered significant interest due to their ability to demonstrate all-optical magnetization reversal [31]. This phenomenon involves the use of ultrashort laser pulses to flip the direction of magnetization upon single or multiple pulse excitation. The mechanism underlying this process relies on ultrafast heating effects and the interplay between the different magnetic properties of TM and RE ions [5, 23]. In our study, we utilized a TbCo alloy with a specific chemical composition of $Tb_{26}Co_{74}$. This alloy is known to be a ferrimagnet, characterized by the presence of two antiparallel magnetic sublattices formed by Tb and Co ions. In this particular stoichiometry, the magnetization of the Tb sublattice surpasses that of the Co sublattice at room temperature, and the magnetic compensation point lies above the room temperature. The TbCo film employed in our experiment exhibits an out-of-plane magnetic anisotropy, contributing to its unique magnetic properties [32]. Further details regarding the sample structure can be found in the Methods section.

The dynamic response of our system was investigated using a conventional pump-probe scheme, as depicted in Fig. 1a. Experiments were performed in an external magnetic field applied perpendicular to the sample plain $H_\perp$=5 kOe. We used ultrashort laser pulses with 300 fs $\lambda$=1030 nm to excite the material while the magnetization change was tracked by analyzing the Kerr rotation of much weaker probe pulses, that impinge the sample with controlled delay with respect to pump pulses. By choosing the probe beam wavelength of either $\lambda$=515 nm or 1030 nm we are able to probe mostly the dynamics of Tb or Co magnetic moments [33, 34, 35]. Further details regarding the magnetization measurement and detection procedure can be found in the Methods and Supplementary information.

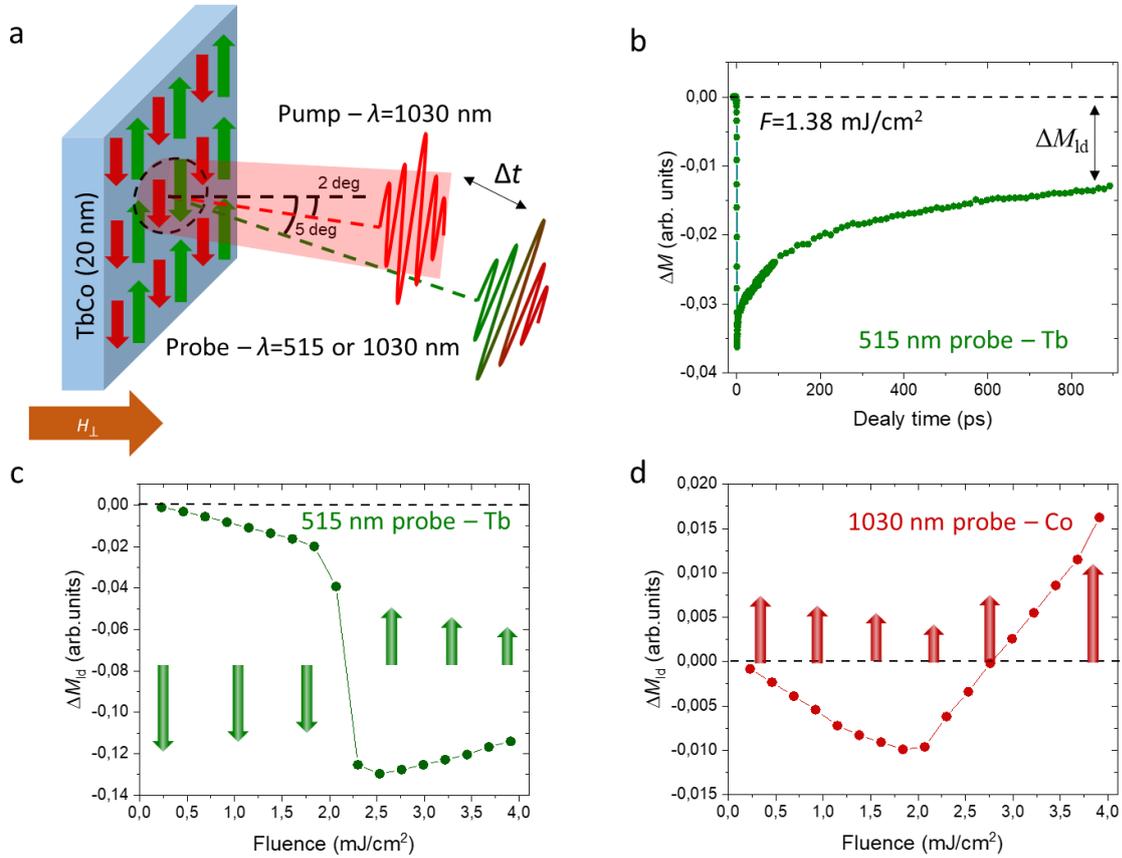

*Figure 1. Optical probe of the magnetization state in TbCo film.* *(a) Experimental geometry for optical pump-probe experiments. (b) Example of time-resolved magnetization dynamics after excitation with 1030 nm pump with fluence F=1.38 mJ/cm² and probed with 515 nm probe. (c) and (d) Transient magnetization change $\Delta M_{ld}$ at $\Delta t$ = 900 ps delay time as a function of pump fluence probed with 515 nm and 1030 nm probe respectively. Green and red arrows on panels (c) and (d) schematically represent the magnetization state of Tb and Co sublattices derived from fluence dependence of $\Delta M_{ld}$ value.*

An example of the time-resolved magnetization dynamics after excitation with a $\lambda$ = 1030 nm pump and fluence of $F$=1.38 mJ/cm² and probed with a $\lambda$ = 515 nm probe is presented in Fig. 1b. The magnetization is initially suppressed on a picosecond timescale, followed by subsequent recovery dynamics. While the obtained time-resolved magnetization trace is similar to previously reported results, there is one notable difference – a transient magnetization change at a long delay time of approximately $\Delta t$ = 900 ps, $\Delta M_{ld}$, which is approximately half of the initially quenched magnetization. This value is unusually large. While the recovery dynamics in magnetic metals can vary significantly, the magnetization state after excitation with light of moderate fluence typically almost fully recovers within hundreds of picoseconds after excitation [36, 37, 38].

To comprehend our results, we monitored the $\Delta M_{ld}$ value as a function of pump fluence probed at $\lambda$=515 nm (Fig. 1c) and 1030 nm (Fig. 1d). The colored arrows in each panel schematically represent the magnetization state of each sublattice at $\Delta t$ = 900 ps after the optical excitation, as determined by $\Delta M_{ld}$ vs. fluence analysis. For a ferrimagnetically coupled two-sublattice system, one would expect similar dependencies regardless of the probing wavelength, as the exchange interaction would equilibrate the magnetization state at such a long delay time. However, we observed substantial differences between the two dependencies. $\Delta M_{ld}$ probed with $\lambda$=515 nm displayed a linear change up to a pump fluence of 2 mJ/cm$^2$, followed by a significant step-like alteration and subsequent linear evolution with an opposite inclination at higher fluences. In contrast, the $\Delta M_{ld}$ dependence probed with $\lambda$=1030 nm did not exhibit a step-like change but showed a change in inclination around 2 mJ/cm$^2$. Remarkably, for pump fluences above 2.73 mJ/cm2, the transient magnetization probed at $\lambda$=1030 nm exceeded the equilibrium value. The nonlinear step-like behavior in Fig. 1c indicates the reversal of the Tb magnetic moment after the optical excitation [39, 40]. Conversely, the absence of a step-like feature in Fig. 1d suggests that the Co magnetic moment's direction remains unchanged, only its amplitude varies. By combining the data from Fig. 1c and Fig. 1d, we conclude that the optical excitation causes a rearrangement of the magnetic moments of Tb and Co from an antiparallel to a parallel configuration. This differs from the all-optical magnetization reversal observed in TM-RE alloys, where both sublattices' magnetization flips. When applying our experimental protocol to the GdFeCo alloy, known for single-shot all-optical magnetization switching [4, 5], and tracking $\Delta M_{ld}$, we observed the characteristic step-like feature in $\Delta M_{ld}$ vs. pump fluence dependence with both probing wavelengths (see SFig. 2)

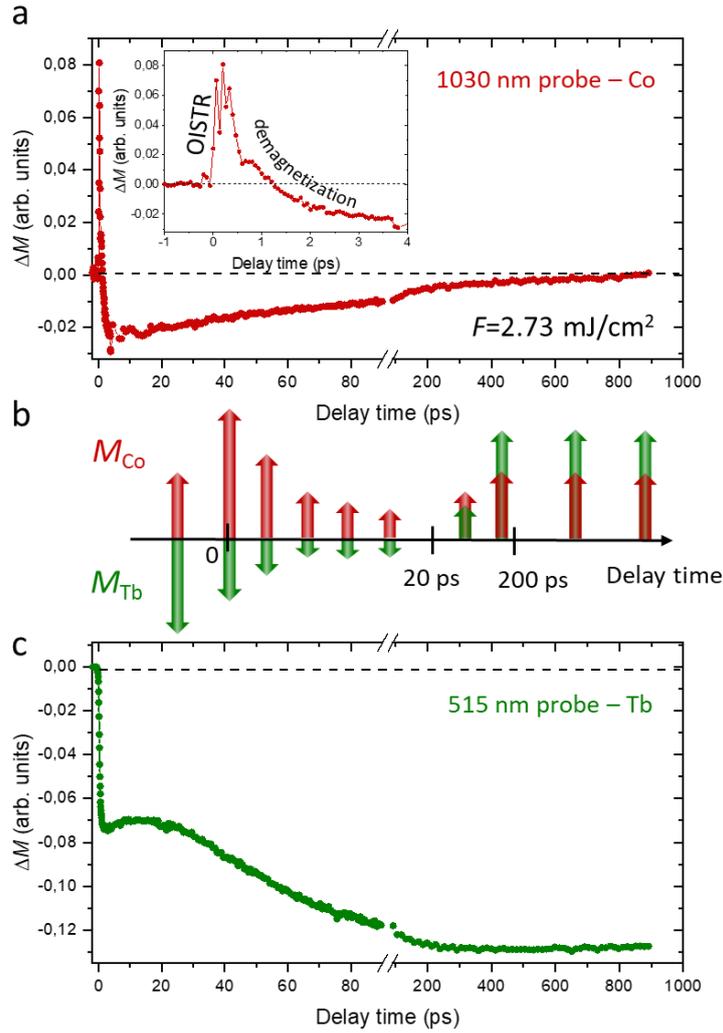

***Figure 2. Time-resolved magnetization dynamics of different sublattices.*** *(a) and (c) Transient magnetization dynamics after excitation with 1030 nm pump with fluence F=2.73 mJ/cm², which is above the threshold for ferri-ferro transition in TbCo, probed with 1030 nm and 515 nm probe, respectively. Panel (b) schematically represents the temporal evolution of the magnetization state of two magnetic sublattices as derived from time-resolved traces in panels (a) and (c).*

Fig. 2 illustrates the time evolution of the ferri- to ferromagnetic transition in TbCo after excitation with a pump pulse of $F=2.73$ mJ/cm², surpassing the threshold of 2 mJ/cm². The dynamics probed with different wavelengths exhibit notable distinctions. For $\lambda = 515$ nm (Fig. 2a), mainly sensitive to Tb moments, the magnetization initially drops within a picosecond, and subsequent recovery dynamics lead to an orientation opposite to equilibrium. In contrast, $\lambda = 1030$ nm (Fig. 2a), predominantly sensing Co moments, displays a rapid magnetization increase within sub-picoseconds, followed by demagnetization and recovery dynamics. At a given excitation fluence, $\Delta M_{ld}$ probed at $\lambda = 1030$ nm is nearly equal to equilibrium, while $\Delta M_{ld}$ probed at $\lambda = 515$ nm significantly differs. The schematic depiction of Tb and Co magnetic moment size and direction evolution derived from the time-resolved traces is presented in Fig. 2b. Notably, at longer delay times, the Tb magnetization surpasses that of Co, as per equilibrium values. Our measurements determine the relative change in optically probed magnetic moment relative to equilibrium for each sublattice but not the absolute magnetic moment value. Based on the data presented in Fig. 2a and 2c, the ferromagnetic coupling in the TbCo alloy is established at around $\Delta t = 20$ ps after the optical excitation, where the flipping of the Tb moments begins, and the system fully reaches a ferromagnetic state at $\Delta t = 200$ ps. Within the $\Delta t = 900$ ps after the excitation, the longest accessible delay time in our experimental setup, we did not observe any indication that the material returns to the ferrimagnetic arrangement. This observation suggests that the ferro-TbCo configuration is metastable and could potentially persist for an extended period.

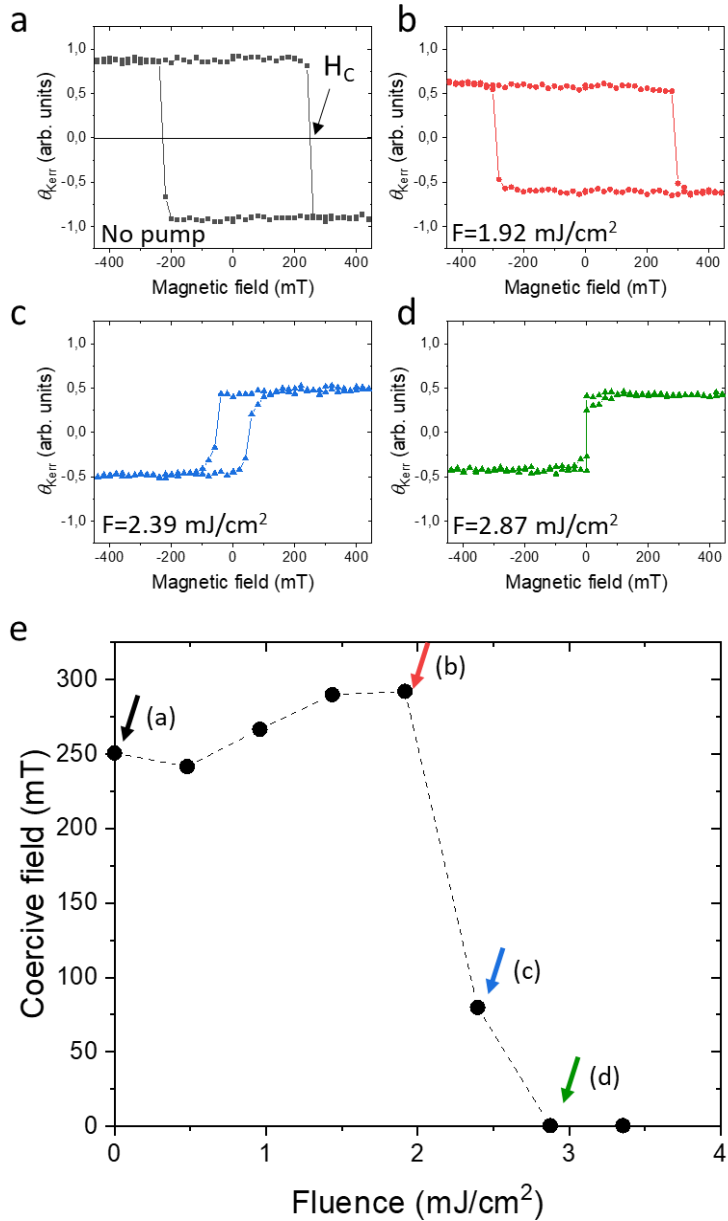

*Figure 3. Magnetic properties of TbCo film when exposed to light.* (a)-(d) magnetic hysteresis loops recorder using magneto-optical Kerr effect with 515 nm probe beam at delay time 900 ps under illumination with pump having different fluence. (e) Coercive field as a function of pump fluence. Colored arrows indicate fluencies during magnetic hysteresis loop measurements, shown in panels (a)-(d).

The experimental findings indicate that TbCo in a ferromagnetic state exhibits different magnetic properties compared to its ferrimagnetic form. Hysteresis loops at $\Delta t = 900$ ps, probed at $\lambda = 515$ nm under varying pump fluences, are shown in Fig. 3a-d. Without pumping, TbCo displays a square hysteresis loop with $H_C$ = 250 mT, typical for out-of-plane anisotropy (Fig. 3a). As pump fluence increases, the coercive field gradually rises, reaching $H_C$ = 300 mT just below the threshold. Transitioning to the ferromagnetic state results in reversed hysteresis loop signs and very low coercivity field values, below what can be resolved with our setup. When applying the same measurement to GdFeCo, hysteresis loops under pump illumination, both below and above the all-optical switching threshold, have identical coercive field values, as expected for magnetization reversal (see SFig. 3). Comparing the hysteresis loops between TbCo and GdFeCo confirms that the changes in TbCo represent a metastable transformation into a ferromagnetic state, distinct from magnetization reversal. This transformation suggests a change in the exchange interaction sign between Tb and Co, resulting in a different magnetic order despite the same chemical composition.

**Discussion**. The exchange interaction in TM-RE alloys, responsible for establishing their magnetic order, is intricately linked to the electronic configuration of the 3*d* states in TM and 5*d* states in RE ions [42, 43]. The sign of the exchange constant between TM and RE depends on the number of vacant positions in each *d*-state, influencing electron hopping between hybridized *d*-sites. The orientation of the highly localized 4*f* moments in RE ions, which significantly contributes to the RE sublattice's magnetic moment, is determined by intra-atomic exchange interactions involving the RE 5*d* and 4*f* states. Consequently, changing the exchange constant's sign between Tb and Co necessitates an optical modification of the electronic configuration of Co 3*d* and Tb 5*d* states that persists for an extended duration.

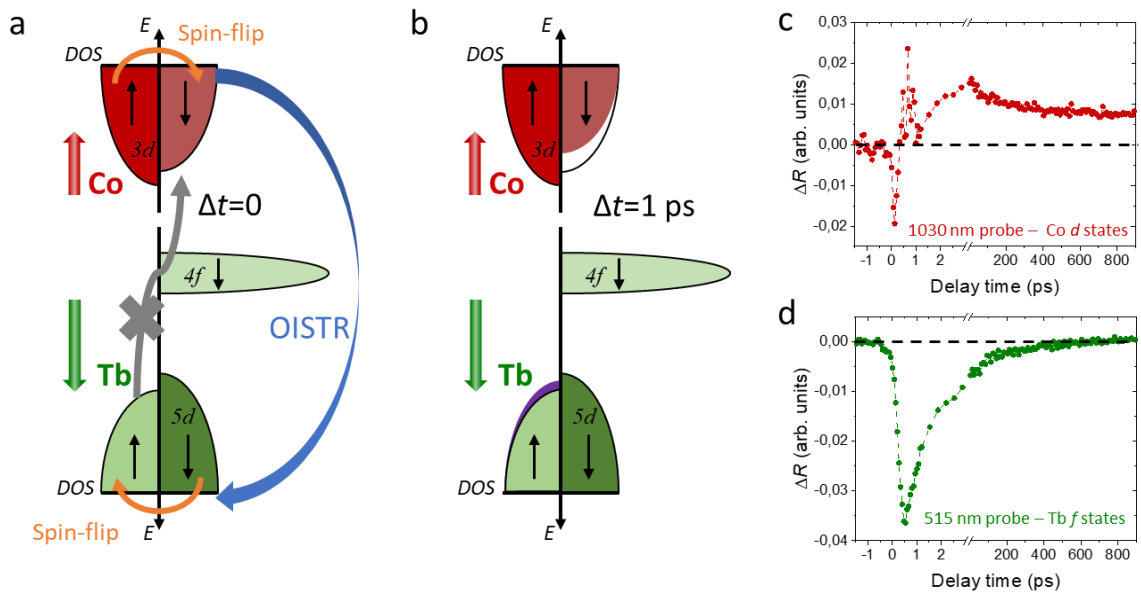

*Figure 4. Electron dynamics in TbCo after the optical excitation*. *(a) Schematic representation of electron traffic routes after the optical excitation and (b) proposed change of the spin accupancy in Co 3d and Tb 5d states indicating that the system does not recover to the equilibrum configuration on short time scale. Proposed state when spins from Co minority band that are trapped in Tb minority band are indicated with violet area. Different resulting occupnasy of Co 3d and Tb 5d states results in altering exchange interaction between Co and Tb sublattices. (c) and (d) transient reflectivity change after excitation with a 1030 nm pump with fluence F=2.73 mJ/cm² probed with 1030 nm and 515 nm light respectively.*

The data in Fig. 2c suggest the occurrence of the Optically Induced Spin-Transfer (OISTR) process, where the magnetic signal in Co specifically increases during interaction with the pump pulse [29, 41]. In OISTR, spins from the minority band of certain ions move to other ions, resulting in a transient magnetization increase as seen in Fig. 2a. OISTR likely occurs between Co $3d$ and Tb $5d$ sites due to their proximity in energy (refer to Fig. 4a and SFig. 4). During OISTR, the spin quantum number is conserved, allowing Co spins to move from the minority band $\mu^{Co}_\uparrow$ to the majority band $\mu^{Tb}_\uparrow$ of Tb $5d$ states, as illustrated by the blue arrow in Fig. 4a. Note that for antiparallel alignment of Tb and Co magnetization, minority and majority bands for two magnetic ions are antiparallel. Simultaneously, optical excitation triggers another process known as spin-flip scattering [30], leading to ultrafast demagnetization (indicated by the orange arrows in Fig. 4a). In this process, spins from the majority band transfer to the minority band within the same ions. Spin-flip demagnetization occurs within a few hundred femtoseconds after optical excitation. Typically, in experiments on ultrafast demagnetization and OISTR, laser pulses with a duration of about 50 fs are used, separating OISTR and spin-flip scattering in time. However, when pumping the system with 300 fs optical pulses, as in our experiment (see Methods for details), both processes occur simultaneously. Consequently, three electron flow processes take place concurrently: i) electrons move from the $\mu^{Co}_\downarrow$ to $\mu^{Co}_\uparrow$ bands in Co (mediated by spin-flip scattering in Co), ii) electrons move from $\mu^{Co}_\uparrow$ to $\mu^{Tb}_\uparrow$ (mediated by OISTR), and iii) electrons move from $\mu^{Tb}_\uparrow$ to $\mu^{Tb}_\downarrow$ band in Tb (mediated by spin-flip scattering in Tb). OISTR is suppressed as the pump pulse leaves the material, and shortly after, the spin-flip process also ceases. A fraction of electrons from Co sites becomes trapped in the $\mu^{Tb}_\downarrow$ sites, as depickted in Fig. 4b. These electrons cannot move back to $\mu^{Co}_\uparrow$ due to spin quantum number restrictions (grey arrow in Fig. 4a). Similarly, they cannot go to $\mu^{Co}_\downarrow$ as this would increase magnetization, which is energetically unfavorable. Consequently, the system ends up in an electronic configuration that cannot quickly return to the ground state. The lack of electrons in the Co minority band explains the increased magnetization state at a long delay time when pumped with a large fluence (see Fig. 1d).

The transient reflectivity signal $\Delta R$ reveals light-induced electronic changes. The dynamics of $\Delta R$ probed at $\lambda = 1030$ nm and 515 nm are shown in Fig. 4c and Fig. 4d, respectively. It's worth noting that with $\lambda = 1030$ nm probe, we primarily track Co $3d$ sites as the amount of Tb $5d$ electrons is smaller (see SFig. 4). The $\Delta R$ trace probed at 515 nm (related to the $4f$ electronic configuration) nearly fully recovers at $\Delta t = 50$ ps after excitation, as expected after optically induced demagnetization processes in magnetic metals [36, 37, 38]. However, the $\Delta R$ trace probed at $\lambda = 1030$ nm (mostly related to the $3d$ electrons in Co) does not fully recover within the accessible delay time range, indicating that the d states' electronic configuration remains optically modified for an extended duration. This observation goes in line with the considerations presented above.

Optically induced magnetic phase transitions are well-documented, often involving transitions from antiferromagnetic states to ferromagnetic or weak ferromagnetic states following optical excitation. However, our case is distinct. Unlike situations where ultrashort pulses drive a system to a known phase reachable by changing temperature, our optical excitation transforms the TbCo alloy into an entirely unprecedented state, one that cannot be achieved through any other means, including temperature changes. It's crucial to distinguish this metastable ferromagnetic order in TbCo from the transient ferromagnetic-like state observed in GdFeCo [23]. In GdFeCo, the transient state lasts for only half a picosecond and is due to unequal demagnetization dynamics of Fe and Gd moments. In contrast, we confirm a ferromagnetic state in TbCo lasting nearly a nanosecond. Although we can't precisely determine the duration of the ferro-TbCo configuration, extrapolating our data suggests it's approximately 50 ns. Given the low threshold fluence for this transition, it might be feasible to maintain TbCo in a ferromagnetic state continuously under pulsed light with a repetition rate of about 20 MHz. This presents a paradigm shift for electronic device design. Until now, electronic components were designed with static roles. However, these results suggest the potential for elements that can dynamically change their functionality under constant ultrashort laser pulse illumination, paving the way for dynamic electronics.

**Achnowlegments**: This work is part of a project which has received funding from the European Union's Horizon 2020 research and innovation program under grant agreement No. 737093, "FEMTOTERABYTE". Sangeeta Sharma would like to thank Leibniz Professorin Program (SAW P118/2021) for funding.

**Author contributions:** SP conceived the project, performed the experiments and analyzed the data. AAL and VK prepared the sample. SP, SS and AS discussed the results. SS made DOS calculations. SP wrote the manuscript with the help of AS and contributions from all co-authors.

**Methods**

**Sample.** The material used in this study is TbCo thin film. An amorphous 20 nm-thick film with a nominal composition of $Tb_{26}Co_{74}$ was sputter deposited with DC magnetron using Tb and Co targets in an argon atmosphere onto a $SiO_2$ substrate. A 2 nm - thick $Al_2O_3$ capping layer was deposited on top of the TbCo film to prevent oxidation. The thin film preparation procedure, together with detailed structural and magnetic characterizations, are presented in detail in Ref. [32]. TbCo is a ferrimagnetic alloy with two antiparallel magnetic sublattices made of Tb and Co. For the given concentration of magnetic elements, the film has an out-of-plane magnetic anisotropy with a coercive field around $H_C$ = 250 mT.

**Time-resolved magnetization dynamics measurements.** Time-resolved pump-probe experimental approach was used to track the evolution of magnetic moments after the light excitation. Ultrashort laser pulses were generated with an ActiveFiber Yb fiber laser with a 50 kHz repetition rate and a fundamental wavelength of $\lambda$ = 1030 nm. The optical excitation was performed using a linearly S-polarized pump beam with a wavelength of $\lambda$ = 1030 nm, and the dynamics were tracked using a linearly P-polarized probe with a wavelength of $\lambda$ = 1030 nm or 515 nm, obtained by frequency doubling using a beta Barium Borate nonlinear crystal. The combination of thin film polarizers and a half-wave plate was used to control the pump power. The pump beam was modulated with an optical chopper with a frequency of $f$=500 Hz. External magnetic field $H_\perp$=5 kOe, generated by NdFeB permanent magnet, was applied perpendicular to the sample surface.

During our time-resolved measurements, we get information about the transient change of the magnetization. To assess the temporal evolution of the magnetization state, we monitored the change in the polarization rotation of the probe beam, which was influenced by the magneto-optical Kerr effect, after the optical excitation affected the magnetization. Our experimental setup specifically detects the transient change of the out-of-plane component of the magnetization. The magneto-optical Kerr effect used in our study to probe the magnetization state is wavelength-dependent. In the case of multisublattice magnetic materials, the resulting Kerr rotation $\theta_K$ is the sum of contributions from each individual sublattice and is proportional to the magnetization of the respective sublattice multiplied by the wavelength-dependent magneto-optical constant. For the TbCo alloy, the measured Kerr rotation can be expressed as follows:

$$\theta_K(\lambda) = K_{Tb}(\lambda)\vec{M}_{Tb} + K_{Co}(\lambda)\vec{M}_{Co}$$

where $K_{Tb(Co)}(\lambda)$ represents the magneto-optical constants and $M_{Tb(Co)}$ represents the magnetization of the Tb and Co sublattices, respectively. In RE-TM alloys, the magneto-optical constant $K$ for the TM sublattice is stronger in the NIR part of the spectrum, while for the RE sublattice, it is stronger in the ultraviolet-visible (UV-VIS) part of the spectrum. This enables us to track the magnetization dynamics of individual sublattices by using probe light with different wavelengths. In our experiments, the magnetization dynamics probed with $\lambda$ = 1030 nm or 515 nm light describe mostly the dynamics in the Co or Tb sublattices, respectively.

Magnetization dynamics, probed with optical methods, might have a parasitic contribution to the signal coming from the transient change of the optical properties of the sample itself. This effect is particularly strong when the pump and probe beam have the same wavelength. In order to get rid of this contribution we performed the experiments with the magnetic field applied in two opposite directions. The resulting transient change of the magnetization is a difference between the traces recorded with two opposite directions of the external magnetic field (see SFig. 1)

**Magnetic hysteresis loop measurements.** During the magnetic hysteresis measurements, the optical modulation with a mechanical chopper was implemented to the probe beam. With this approach, we were able the measure the absolute Kerr rotation of the probe beam. The magnetic field during the hysteresis measurements was changed by moving the permanent magnet in the direction perpendicular to the sample surface. We would like to note that the approach with modulation of the probe beam could be used for time-resolved measurements, but it suffers from a much worse signal-to-noise ratio compared with the method that was described above for time-resolved dynamics.

**Densuty of states calculations**. Theoretically, the density of states (DOS) is calculated by first performing a ground-state calculation using density functional theory (DFT) within the local spin density approximation

(LSDA) for the exchange-correlation potential [45]. A Hubbard $U = 6.7$ eV was used to treat localized *f*-states of the Tb atom [46]. Followed by this, atom and angular momentum projected DOS were calculated by spherical harmonic expansion of the density within a sphere around each atom. All these calculations are performed using the state-of-the art full-potential linearized augmented plane wave method as implemented in the Elk code [47]. In order to obtain converged results, a k-point grid of 12x12x12 was used. All states greater than 95 eV below Fermi-level were treated as Dirac spinors, i.e. obtained by solving the Dirac equation. All the other states are treated as Pauli spinors obtained by solving the Schrödinger equation including spin-orbit and other relativistic corrections (e.g mass correction and Darwin terms).

# Supplementary materials

1. Optical probe of the magnetization

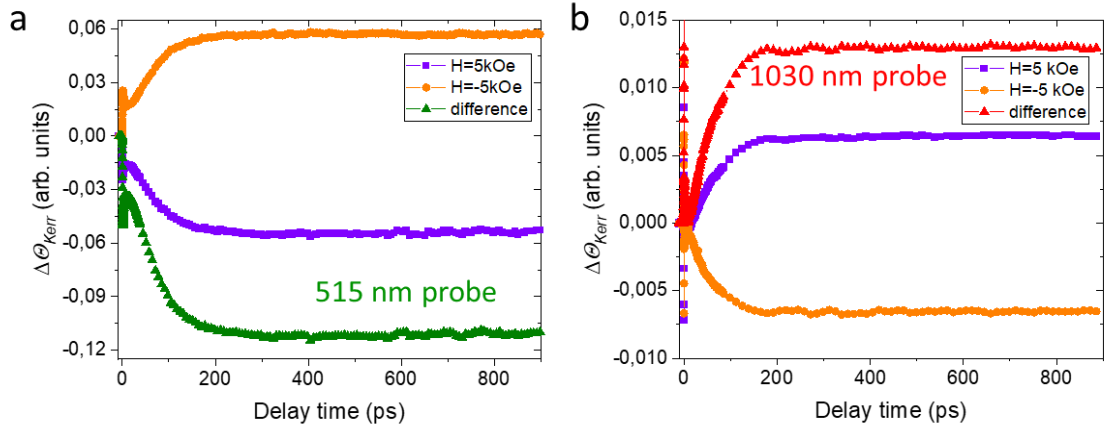

*SFigure 1. Time-resolved Kerr rotation of probe beam with λ=515 nm (a) and λ=1030 nm (b) for opposite sign of the external magnetic field and respective difference after excitation with F=3.91 mJ/cm² pump in TbCo alloy.*

The magneto-optical Kerr effect allows us to get information about the magnetization state of the material by analyzing the change in rotation of the polarization plane of linearly polarized light reflected from the magnetic material. However, the obtained information about the magnetization might be wrong if the reflectivity of the sample is changing. This is the case during time-resolved magnetization measurements and is especially strong when the pump and probe beams have the same wavelength. However, the parasitic contribution could be significantly reduced if the measurements were done with two opposite directions of the external magnetic field and the difference between two traces is taken. SFig. 1 shows the magnetization dynamics in TbCo after excitation with a pump beam having fluence of $F=3.91$ mJ/cm² probed with $\lambda=515$ nm and $\lambda=1030$ nm probe for two opposite orientations of the external magnetic field. The transient magnetization signal $\Delta M$, discussed in the main text, is

$$\Delta M = \theta_{Kerr}^{H+} - \theta_{Kerr}^{H-}$$

With this method, we are able to significantly suppress artefacts that come from transient reflectivity change in our experiments.

2. All optical magnetization switching in GdFeCo probed with $\lambda$=515 nm and $\lambda$=1030 nm

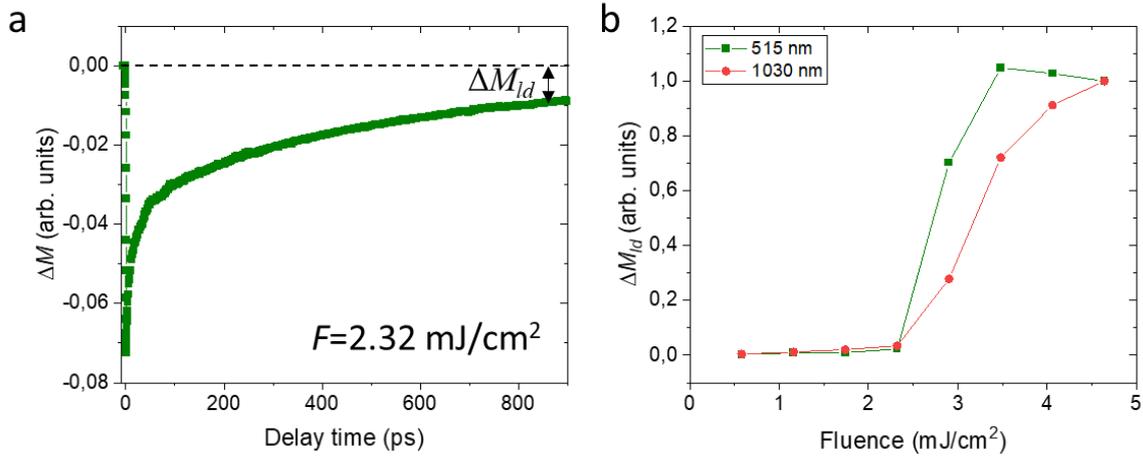

*SFigure 2. (a) Example of time-resolved magnetization dynamics after excitation with $\lambda$=1030 nm pump with fluence F=2.32 mJ/cm² and probed with $\lambda$=515 nm probe. (b) Transient magnetization change $\Delta M_{ld}$ at $\Delta t$ = 900 ps long delay time as a function of pump fluence probed with $\lambda$=515 nm and $\lambda$=1030 nm probe respectively. Curves are normalized to their value at F=4.64 mJ/cm².*

SFigure 2 shows an example of time-resolved magnetization dynamics in $Gd_{27}(Fe, Co)_{73}$ alloy taken under the same excitation scheme as the results on TbCo presented in the main text. This material is known to exhibit single-shot all-optical magnetization switching. Transient change of the magnetization at $\Delta t$ = 900 ps long delay $\Delta M_{ld}$ (see SFig. 2a) was monitored as a function of pump fluence. Probing the magnetization with both $\lambda$=515 nm and $\lambda$=1030 nm light we observed a step-like feature in $\Delta M_{ld}$ vs. fluence dependence, indicating the reversal of both RE and TM sublattices. Such behavior is expected in the case of regular all-optical magnetization switching. Some deviation in inclination is most likely due to a small mismatch in the beam size for $\lambda$=515 nm and $\lambda$=1030 nm probe beams.

3. Magnetic hysteresis loops for GdFeCo alloy under the pump illumination

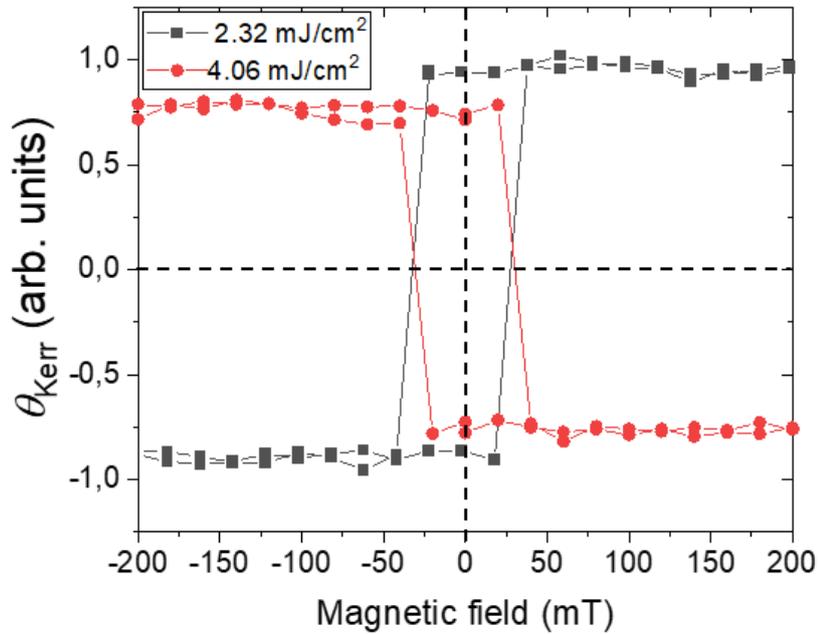

*SFigure 3. Magnetic hysteresis loops in GdFeCo alloy under the pump illumination with fluencies above and below the AOS fluence threshold. When illuminated with a pump having fluence above the threshold for AOS, the magnetic hysteresis loop changes the sign but coercive field $H_C$ remains the same as when illuminated with a pump having the fluence below the AOS threshold. This is an expected result in the case of regular magnetization reversal as the magnetic properties of the material remain the same unlike in the case for TbCo, described in the main text.*

4. Density of states calculation

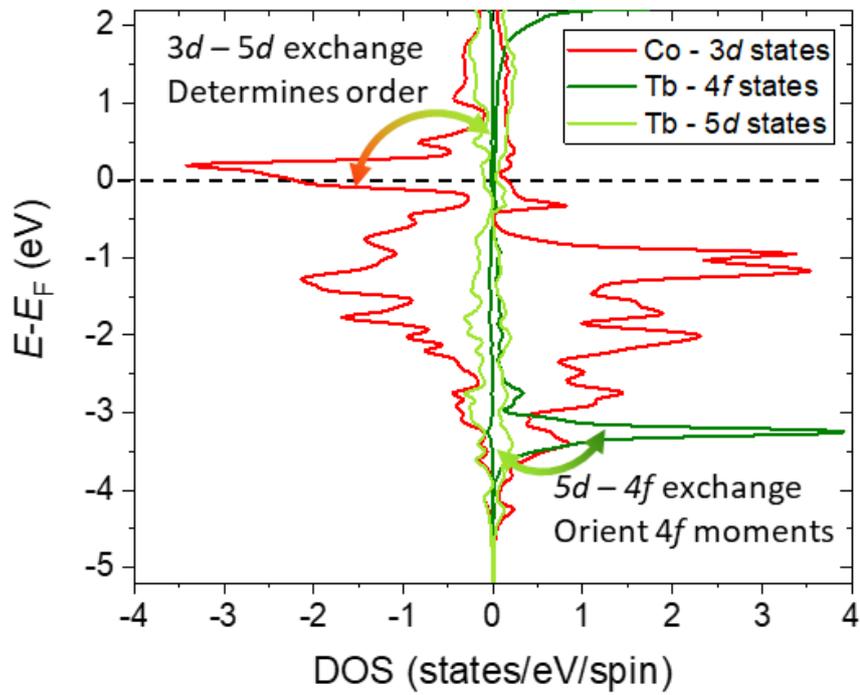

*SFigure 4. Density of state (DOS) calculation for Co 3d-states and Tb 5d- and 4f-states. Tb 4f-states are localized around 3.5 eV below $E_F$. Arrows schematically show different types of exchange interactions in TbCo alloy.*